\documentclass[aps,prb,twocolumn,showpacs,amsmath]{revtex4}
\usepackage{CJK}
\usepackage{graphicx}
\usepackage{epstopdf}
\usepackage{amsmath}
\usepackage{indentfirst}
\usepackage{float}
\usepackage{subfigure}
\usepackage{natbib}

\begin{document}

\title{Enhanced negative nonlocal conductance in an interacting quantum dot connected to two ferromagnetic leads and one %%@
superconducting lead}

\author{C. Lee}
\affiliation{Key Laboratory of Artificial Structures and Quantum Control (Ministry of Education), Department of Physics and %%@
Astronomy, Shanghai Jiaotong University, 800 Dongchuan Road, Shanghai 200240, China}

\author{Bing Dong}
\thanks{Author to whom correspondence should be addressed. Email:bdong@sjtu.edu.cn.}
%\email{bdong@sjtu.edu.cn.}
\affiliation{Key Laboratory of Artificial Structures and Quantum Control (Ministry of Education), Department of Physics and %%@
Astronomy, Shanghai Jiaotong University, 800 Dongchuan Road, Shanghai 200240, China}
\affiliation{Collaborative Innovation Center of Advanced Microstructures, Nanjing, China}

\author{X. L. Lei}
\affiliation{Key Laboratory of Artificial Structures and Quantum Control (Ministry of Education), Department of Physics and %%@
Astronomy, Shanghai Jiaotong University, 800 Dongchuan Road, Shanghai 200240, China}
\affiliation{Collaborative Innovation Center of Advanced Microstructures, Nanjing, China}

\begin{abstract}

In this paper, we investigate the electronic transport properties of a quantum dot (QD) connected to two ferromagnetic leads and %%@
one superconductor lead in the Kondo regime by means of the finite-$U$ slave boson mean field approach and nonequilibrium Green %%@
function technique. In this three-terminal hybrid nano-device, we will focus our attention on the joint effects of the Konod %%@
correlation, superconducting proximity pairing, and spin polarization of leads. It is found that: the superconducting proximity %%@
effect will suppress the linear local conductance (LLC) stemming from the weakened Kondo peak, and when its coupling $\Gamma_s$ %%@
is bigger than the tunnel-coupling $\Gamma$ of two normal leads, the linear cross conductance (LCC) becomes negative in the Kondo %%@
region; for antiparallel configuration, increasing spin polarization further suppresses LLC but enhances LCC, i.e. causing larger %%@
negative values of LCC, since it is benefit for emergence of cross Andreev reflection; On the contrary, for parallel %%@
configuration, with increasing spin polarization, the LLC descends and greatly widens with the appearance of shoulders, and %%@
eventually splits into four peaks, and meanwhile the LCC reduces relatively rapidly to the normal conductance.

\end{abstract}


\maketitle

\section{Introduction}

Recently, electron transport through hybrid nanodevice, for instance, a quantum dot (QD) connected to normal and  superconducting %%@
electrodes, has attracted much attention in many experimental %%@
\cite{PhysRevLett.89.256801,PhysRevLett.95.027002,PhysRevLett.97.237003,jarillo2006quantum,van2006supercurrent,PhysRevLett.99.126%%@
603,hofstetter2009,PhysRevB.79.134518,franceschi2010,Franke940,dirks2011quantum,PhysRevLett.109.157002,PhysRevLett.111.136806,lee%%@
2014quantum,deacon2015quantum,PhysRevLett.115.087001,PhysRevLett.115.227003,albrecht2016,PhysRevLett.117.186801,PhysRevLett.105.0%%@
77001} and theoretical %%@
studies,\cite{PhysRevB.63.094515,PhysRevB.82.184507,PhysRevB.82.134508,doi:10.1080/00018732.2011.624266,PhysRevB.84.075484,PhysRe%%@
vB.85.094518,PhysRevB.87.115409,PhysRevB.92.235422,PhysRevB.93.195125} due to their physical challenges and potential %%@
applications in spintronics and quantum information. When a QD is connected to a superconductor, superconducting order can leak %%@
into it to give rise to pairing correlations and an induced superconducting gap, known as the superconducting proximity effect, %%@
which privileges the tunnelling of Cooper pairs of electrons with opposite spin, and thereby favours QD states with even numbers %%@
of electrons and zero total spin. At the same time, the local Coulomb repulsion enforces a one-by-one filling of the QD, and %%@
thereby induces the Coulomb blockade and even the Kondo effect at considerably low temperature, which exhibits zero-bias anomaly %%@
in the differential conductance with odd number of electrons residing in the QD.
In this case, the superconducting proximity effect competes with the on-site Coulomb %%@
correlation.\cite{PhysRevLett.89.256801,Franke940,PhysRevLett.99.126603,PhysRevB.63.094515,PhysRevB.92.235422,doi:10.1080/0001873%%@
2.2011.624266,PhysRevB.84.075484,PhysRevB.93.195125} 

It is even more intriguing when the QD additionally connects to a ferromagnetic %%@
lead.\cite{PhysRevLett.93.197003,PhysRevLett.104.246804} It has been already known that, the effective exchange field induced by %%@
the ferromagnetic correlation can cause spin imbalance inside the QD, and as a result, suppress and/or even split the Kondo peak %%@
in the differential conductance.\cite{PhysRevLett.91.127203,Pasupathy86,hauptmann2008,PhysRevLett.107.176808,Dong_2003} Besides, %%@
spin polarization of the QD, on the one hand, is disadvantageous to the formation of on-dot superconducting pairing. But the spin %%@
polarization in the antiparallel configuration, on the other hand, is favorable to the Adnreev reflection (AR) and Cooper pair %%@
splitting.\cite{PhysRevLett.93.197003} It is, therefore, very interesting to study how the interplay of the Kondo, %%@
superconducting pairing, and ferromagnetic correlations affects the electron tunneling through a QD.\cite{PhysRevB.92.245307} In %%@
a recent paper, D. Futterer {\it et al} present a theoretical analysis of the subgap transport of such a three-terminal hybrid %%@
system, consist of an interacting QD attached to two ferromagnetic leads and one superconducting %%@
lead.\cite{PhysRevB.79.054505,Futterer2010} They focused on the first-order sequential tunneling by using master equation and %%@
found that the strong on-dot electron-electron interaction, rather than the nonlocal AR, leads to negative values of the nonlocal %%@
current response at an appropriately large bias voltage. 
Moreover, the tunneling magnetoresistance has been calculated for the same system to display a nontrivial dependence on the bias %%@
voltage and the level detuning due to the AR.\cite{PhysRevB.89.115305} 
Later on, it has been nevertheless reported that the cross AR is indeed the dominant nonlocal transport channel at low bias %%@
voltage and leads to a negative value of the cross conductance in the three terminal hybrid nanodevice with two normal electrodes %%@
instead.\cite{PhysRevB.88.155425,srep14572}  

In the present work, we extend a finite-$U$ slave boson mean field (SBMF) approach of Kotliar and %%@
Ruckenstein\cite{PhysRevLett.57.1362} with help of the nonequilibrium Green function (NGF) method to investigate the subgap %%@
transport for the same three-terminal hybrid QD as in Ref.~\onlinecite{PhysRevB.79.054505}.
This kind of SBMF approach is generally believed to be reliable in describing not only spin fluctuations rigorously but also %%@
charge fluctuations to certain degree in the Kondo regime at zero temperature. This nonperturbative approach has been %%@
successfully utilized to calculate the linear and nonlinear conductance within a relatively wide dot-level range from the mixed %%@
valence to the empty orbital regimes, in which the major characteristics induced by the external magnetic field and the %%@
magnetization in Kondo transport arise.\cite{PhysRevB.63.235306,Dong_2001,PhysRevB.65.241304,Jing_2005} Besides, this approach %%@
has been furthermore applied to analyze the $\pi$-phase transition in a double-QDs Josephson junction due to competition between %%@
Kondo and interdot antiferromagnetic coupling.\cite{PhysRevB.74.132505} The main purpose of this paper is thereby to analyze in %%@
detail the interplay of the Kondo, superconducting proximity induced on-dot pairing, and ferromagnetic correlations and their %%@
influence on electronic tunneling.  

The rest of the paper is organized as follows. In Sec.II, we introduce our model of the three-terminal hybrid system, and
the equivalent slave-boson field Hamiltonian. Then we present the self-consistent equations of the expectation values of %%@
slave-boson operators within the SBMF approach and NGF method. Moreover, the formulas for current and linear conductance, %%@
including the local and cross conductances, are given. In Sec.III, we present and analyze in detail our numerical calculations %%@
for the linear conductance and nonlinear conductance. Finally, a brief summary is given in Sec.IV.

\section{Model and theoretical formulation}

\subsection{Model Hamiltonian}

\begin{figure}[htb]
\includegraphics[height=4.5cm,width=8cm]{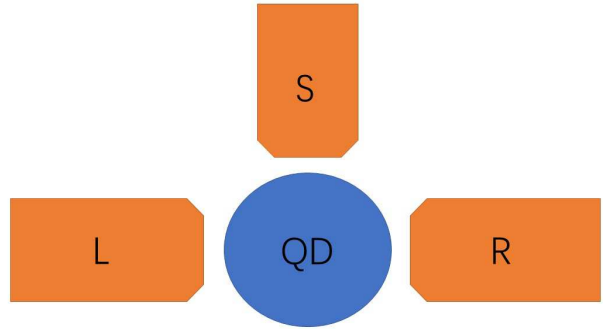}
\caption{(Colour online) Schematic diagram of a quantum dot connected to one superconducting lead and two ferromagnetic leads.}
\label{fig1}
\end{figure}

We consider a three-terminal hybrid nanodevice: an interaction QD connected to one superconducting lead and two ferromagnetic %%@
leads, as shown in Fig.~\ref{fig1}.
The Hamiltonian of the system can be written as:\cite{PhysRevB.79.054505}
\begin{equation}\label{1}
H=H_{L}+H_R+H_{QD}+H_{T},
\end{equation}
where
\begin{equation}\label{2}
H_{\eta}=\sum_{k\sigma}\epsilon_{\eta k\sigma}c_{\eta k\sigma}^{\dag}c_{\eta k\sigma},
\end{equation}
\begin{equation}\label{3}
H_{QD}=\sum_{\sigma} \epsilon_{d} c_{d\sigma}^{\dag}c_{d\sigma}+Un_{1}n_{2}+\Gamma_{s}
(c_{d1}^{\dag}c_{d2}^{\dag}+c_{d1}c_{d2}),
\end{equation}
\begin{equation}\label{4}
H_{T}=\sum_{\eta k\sigma}\left ( V_{\eta k}c_{\eta k\sigma}^{\dag}c_{d\sigma}+ {\rm H.c.} \right ).
\end{equation}
Here $\eta=L,R$ denotes the left and right leads, while $\sigma=1,2$ represents the spin degree of freedom. In the above %%@
equations, $c_{\eta k\sigma}^{\dag}$ ($c_{\eta k\sigma}$) and $c_{d\sigma}^{\dag}$ ($c_{d\sigma}$) are creation (annihilation) %%@
operators of electrons with spin $\sigma$ in the $\eta$th ferromagnetic lead and in the QD respectively. In the dot Hamiltonian  %%@
$H_{QD}$, $\epsilon_{d}$ is the energy level of the QD, $n_{\sigma}=c_{d\sigma}^{\dag}c_{d\sigma}$, and $U$ is the on-site %%@
Coulomb repulsion between opposite spin electrons. $H_T$ depicts the tunneling between the QD and the two ferromagnetic leads, %%@
and $V_{\eta k}$ is the corresponding tunneling matrix element. In general, the tunneling amplitude $V_{\eta k}$ is assumed to be %%@
independent of spin and energy, and thus the effect of spin-polarized tunneling is captured by the spin-dependent tunneling %%@
rates, $\Gamma_{\eta\sigma}=2\pi \sum_{k}|V_{\eta k}|^2 \delta(\omega-\epsilon_{\eta k \sigma})$.

In this paper, since we are only interested in the subgap tunneling, it is natural to consider the limit of extremely large %%@
superconducting gap in the superconducting lead. Therefore, the degree of freedom of the superconducting lead can be integrated %%@
out and an effective term can be constructed in the dot Hamiltonian, the third term in Eq.~(\ref{3}). The parameter $\Gamma_s$ %%@
plays indeed the role of describing the superconducting proximity effect on the dot. It is evident that this new proximized term %%@
mixes the empty state $|0\rangle$ and the doubly occupied state $|\uparrow\downarrow\rangle$ in the dot, and results in two new %%@
eigenstates with energies, $E_{\pm}=\varepsilon\pm \sqrt{\varepsilon^2+4\Gamma_s^2}$ (here $\varepsilon=\epsilon_d+U/2$), which %%@
are known as the Andreev bound states. What we are interested in this paper is the effect of Andreev reflection on the electron %%@
tunneling through an interacting QD in the Kondo regime.

According to the finite-$U$ slave-boson approach, one can introduce additional four auxiliary boson operators $e$, $p_\sigma$, %%@
and $d$, which are associated respectively with the empty, singly occupied, and doubly occupied electron states of the QD, to %%@
discuss the above problem without interparticle couplings in an enlarged space with constraints: the completeness %%@
relation,\cite{PhysRevLett.57.1362}
\begin{equation}\label{5}
\sum_{\sigma} p_{\sigma}^{\dag}p_{\sigma}+ e^{\dag}e+ d^{\dag}d =1,
\end{equation}
and the particle number conservation condition,
\begin{equation}\label{6}
c_{d\sigma}^{\dag}c_{d\sigma}= p_{\sigma}^{\dag}p_{\sigma}+ d^{\dag}d.
\end{equation}
Within the mean-field scheme, the effective Hamiltonian becomes:\cite{PhysRevLett.57.1362}
\begin{widetext}
\begin{equation}\label{Ham}
\begin{split}
H=&\sum_{\sigma} \epsilon_{d} c_{d\sigma}^{\dag} c_{d\sigma}+ U d^{\dag}d + \Gamma_{s} %%@
(z_{1}^{*}z_{2}^{*}c_{d1}^{\dag}c_{d2}^{\dag} + z_1 z_2 c_{d1} c_{d2})
+ \sum_{\eta k\sigma}(V_{\eta k}c_{\eta k\sigma}^{\dag}c_{d\sigma}z_{\sigma}
+V_{\eta k}^{*}c_{d\sigma}^{\dag}c_{\eta k\sigma}z_{\sigma}^{*})\\
&+\sum_{\eta k\sigma}\epsilon_{\eta k\sigma}c_{\eta k\sigma}^{\dag}c_{\eta k\sigma}
+\lambda^{1}(\sum_{\sigma}p_{\sigma}^{\dag}p_{\sigma}+e^{\dag}e+d^{\dag}d-1)
+ \sum_{\sigma}\lambda_{\sigma}^{2}(c_{d\sigma}^{\dag}c_{d\sigma}-p_{\sigma}^{\dag}p_{\sigma}-d^{\dag}d),
\end{split}
\end{equation}
where three Lagrange multipliers $\lambda^{1}$ and $\lambda_{\sigma}^{2}$ are drawn in order to make the constraints valid, and %%@
$z_{\sigma}$ is the correctional parameters in the hopping term to recover the many-body effect on tunneling with
\begin{equation}\label{8}
z_{\sigma}=(1-d^\dagger d- p_{\sigma}^\dagger p_{\sigma})^{-1/2} (e^{\dag}p_{\sigma}+p^{\dag}_{\bar{\sigma}}d) (1-e^\dagger e- %%@
p_{\sigma}^\dagger p_{\sigma})^{-1/2}.
\end{equation}

\subsection{Self-consistent equations}

From the effective Hamiltonian Eq.~(\ref{Ham}), one can derive four equations of motion of slave-boson operators, which serve as %%@
the basic equations together with the three constraints. Then we apply further the mean-field approximation in the statistical %%@
expectations of these equations, where all the boson operators are replaced by their respective expectation values. After a %%@
lengthy and tedious calculation by employing Langreth technique, we can obtain the self-consistent equations as %%@
follows:\cite{PhysRevB.63.235306,Dong_2001,PhysRevB.65.241304,Jing_2005}
\begin{equation}
\Gamma_{s} \frac{\partial (z_{1}z_{2})}{\partial e}(R+R^{*})+2\lambda^{1}e+\sum_{\sigma}p_{\sigma}(Q_{\sigma}+Q_{\sigma}^{*})=0,
\end{equation}
\begin{equation}
\Gamma_{s} \frac{\partial (z_{1}z_{2})}{\partial p_1} (R+R^*)+2(\lambda^1-\lambda_1^2)p_1+e(Q_1+Q_1^*)+d(Q_2+Q_2^*)=0,
\end{equation}
\begin{equation}
\Gamma_s \frac{\partial (z_{1}z_{2})}{\partial p_2} (R+R^*)+2 (\lambda^1p_2-\lambda_2^2) p_2+d(Q_1+Q_1^*)+e(Q_2+Q_2^*)=0,
\end{equation}
\begin{equation}
\Gamma_s \frac{\partial (z_{1}z_{2})}{\partial d} (R+R^*) +2(U+\lambda^1d-\sum_{\sigma}\lambda_{\sigma}^2) d + %%@
\sum_{\sigma}p_{\bar{\sigma}}(Q_{\sigma}+Q_{\sigma}^*)=0,
\end{equation}
\begin{equation}
\sum_{\sigma}|p_{\sigma}|^2+|e|^2+|d|^2-1=0,
\end{equation}
\begin{equation}
K_{\sigma}-|p_{\sigma}|^2-|d|^2=0,
\end{equation}
where
\begin{equation}
K_1=\frac{1}{2\pi i}\int d\omega G_{d11}^{<}(\omega),
\end{equation}
\begin{equation}
K_2=\frac{-1}{2\pi i}\int d\omega G_{d22}^{<}(\omega),
\end{equation}
\begin{equation}
R=\frac{1}{2\pi i}\int d\omega G_{d21}^{<}(\omega),
\end{equation}
\begin{equation}
\begin{split}
Q_{1\eta}=& z_1\Gamma_{\eta1} \int \frac{d\omega}{2\pi} \left \{ - \frac{i}{2} \left [ \widetilde{\Gamma}_{L1} f_L(\omega) + %%@
\widetilde{\Gamma}_{R1} f_R(\omega) \right ] |G_{d11}^R(\omega)|^2 \right. \\
& \left. - \frac{i}{2} \left [ \widetilde{\Gamma}_{L2} (1-f_L(-\omega)) + \widetilde{\Gamma}_{R1} (1-f_R(-\omega)) \right ] %%@
|G_{d21}^R(\omega)|^2 + f_{\eta}(\omega)G_{d11}^A(\omega) \right \},
\end{split}
\end{equation}
\begin{equation}
\begin{split}
Q_{2\eta}=& z_2\Gamma_{\eta2} \int \frac{d\omega}{2\pi} \left \{ \frac{i}{2} \left [ \widetilde{\Gamma}_{L1} (1-f_L(\omega)) + %%@
\widetilde{\Gamma}_{R1} (1-f_R(\omega)) \right ] |G_{d21}^R(\omega)|^2 \right. \\
& \left. + \frac{i}{2} \left [\widetilde{\Gamma}_{L2} f_L(-\omega) + \widetilde{\Gamma}_{R1} f_R(-\omega) \right ] %%@
|G_{d22}^R(\omega)|^2 - f_{\eta}(-\omega) G_{d22}^A(\omega) \right \},
\end{split}
\end{equation}
and
\begin{equation}
Q_{\sigma}=\sum_{\eta}Q_{\sigma\eta}.
\end{equation}
Here the QD Keldysh NGFs, $G_{d\sigma\sigma'}^{R(A,<)}(\omega)$ are the matrix elements of the $2\times 2$ retarded (advanced and %%@
correlation) GF matrix $G_{d}^{R(A,<)}(\omega)=\langle\langle \phi; \phi^\dagger \rangle\rangle^{R(A,<)}$ defined in the Nambu %%@
presentation, in which the mixture Fermion operator, $\phi=(c_{d1},c_{d2}^{\dag})^T$,
has to be introduced to describe electronic dynamics due to the superconducting proximity effect.
For the effective noninteracting Hamiltonian, the retarded and advanced GFs $G_d^{R(A)}$ can be easily written in the frequency %%@
domain as:
\begin{equation}
\left( G^{R(A)}(\omega) \right )^{-1} =\left[
\begin{array}{cc}
\omega-\epsilon_d - \lambda_1^2 \pm \frac{i}{2}(\widetilde{\Gamma}_{L1}+ \widetilde{\Gamma}_{R1}) & - \Gamma_{s}z_1z_2\\
-\Gamma_{s} z_1^*z_2^* & \omega + \epsilon_d + \lambda_2^2 \pm \frac{i}{2}(\widetilde{\Gamma}_{L2}+ \widetilde{\Gamma}_{R2})
\end{array}
\right].
\end{equation}
\end{widetext}
with the renormalized parameters, $\widetilde{\Gamma}_{\eta \sigma}=|z_\sigma|^2 \Gamma_{\eta\sigma}$.
And the correlation GF $G_d^{<}(\omega)$ can be obtained with the help of the following Keldysh relation typical for a %%@
noninteracting system,
\begin{equation}
G_d^{<}(\omega)=G_d^{R}(\omega) \Sigma_\phi^<(\omega) G_d^{A}(\omega),
\end{equation}
with the self-energy,
\begin{equation}
\Sigma_\phi^<(\omega) = \sum_{\eta} 2\pi i \left[
\begin{array}{cc}
\widetilde{\Gamma}_{\eta 1} f_{\eta}(\omega) & 0\\
0 & \widetilde{\Gamma}_{\eta 2} [1 - f_{\eta}(-\omega)]
\end{array}
\right],
\end{equation}
where $f_{\eta}(\omega)=1/(e^{\beta(\omega - \mu_\eta)}+1)$ is the Fermi distribution function of the lead $\eta$ with the %%@
chemical potential $\mu_\eta$ and temperature $1/\beta$.

\subsection{The current and linear conductance}

The electric current flowing from the lead $\eta$ into the QD can be obtained from the rate of change of electron number operator %%@
of the left lead,
\begin{equation}
I_\eta=\sum_{\sigma} I_{\eta\sigma}= -e\sum_{\sigma} \left\langle \frac{d}{dt} \sum_{k} c_{\eta k \sigma}^\dagger c_{\eta k %%@
\sigma} \right\rangle.
\end{equation}
After standard calculation, the current for the left lead can be written as\cite{PhysRevB.88.155425,srep14572}
\begin{equation}\label{iL}
I_L=I_L^{ET}+ I_L^{DAR} + I_L^{CAR},
\end{equation}
with
\begin{equation}\label{iet}
\begin{split}
I_L^{ET}=&\frac{e}{h}\int d\omega \left \{ \widetilde{\Gamma}_{L1} \widetilde{\Gamma}_{R1} \left [ f_L(\omega)-f_R(\omega) %%@
\right] |G_{d11}^R(\omega)|^2 \right. \\
& \left. +\widetilde{\Gamma}_{L2} \widetilde{\Gamma}_{R2} \left [ f_L(-\omega)-f_R(-\omega) \right ] |G_{d22}^R(\omega)|^2 \right %%@
\},
\end{split}
\end{equation}
\begin{equation}
\begin{split}
I_L^{DAR}=&\frac{2e}{h}\int d\omega \widetilde{\Gamma}_{L1} \widetilde{\Gamma}_{L2} \left [ f_L(\omega)+f_L(-\omega)-1 \right ] %%@
\\
& \times |G_{d12}^R(\omega)|^2 ,
\end{split}
\end{equation}
\begin{equation}\label{icar}
\begin{split}
I_L^{CAR}=&\frac{e}{h}\int d\omega \left \{ \widetilde{\Gamma}_{L1} \widetilde{\Gamma}_{R2} \left [ f_L(\omega)+f_R(-\omega)-1 %%@
\right ] \right. \\
&\left. +\widetilde{\Gamma}_{L2} \widetilde{\Gamma}_{R1} \left [ f_L(-\omega)+f_R(\omega)-1 \right ] \right \} %%@
|G_{d12}^R(\omega)|^2.
\end{split}
\end{equation}
The corresponding currents for the right lead can be readily obtained by simply exchanging the subscripts L and R in %%@
Eqs.~(\ref{iet})-(\ref{icar}).
It is found that the current can be divided into three parts: $I_L^{ET}$ describes the single-particle tunneling current due to %%@
the normal electron transfer (ET) processes from the left lead directly to the right lead; $I_L^{DAR}$ denotes the local Andreev %%@
current due to the direct AR (DAR) processes in which an electron injecting from the left lead forms a Cooper pair in the %%@
superconducting lead and, at the same time, is reflected as a hole back into the left lead; while $I_L^{CAR}$ is the nonlocal %%@
Andreev current due to the crossed AR (CAR) processes which is similar to DAR except that hole is reflected into another lead, %%@
i.e., here the right lead.

Since we are interested in the interplay between the Andreev bound state and the Kondo effect in the nonlocal subgap tunneling, %%@
we choose the bias voltage configuration in this hybrid three-terminal nanodevice as follows: the left lead is biased with the %%@
chemical potential $V$, while the right lead and the superconducting electrode are both in contact with ground. Therefore, one %%@
can define two different linear conductances, the usual local conductance $G_L=\partial I_{L}/\partial V |_{V=0}$ and the unusual %%@
nonlocal (cross) conductance $G_C=\partial I_{R}/\partial V |_{V=0}$, which is related to the nonlocal current response of the %%@
hybrid three-terminal nanodevice to external driving field, i.e., current flowing in the right lead due to the bias voltage %%@
applied to the left lead. From Eqs.~(\ref{iet})-(\ref{icar}), the local conductance reads
\begin{equation}
G_{L}=\frac{\partial I_{L}}{\partial V}{\bigg |}_{V=0}=G^{ET}+G^{DAR}+G^{CAR},
\end{equation}
and the cross conductance is
\begin{equation}\label{gc}
G_{C}=\frac{\partial I_{R}}{\partial V}{\bigg |}_{V=0}=G^{ET}-G^{CAR},
\end{equation}
where
\begin{equation}
G^{ET}=\frac{e^2}{h}\left ( \widetilde{\Gamma}_{L1} \widetilde{\Gamma}_{R1}
|G_{d11}^R(0)|^2 +\widetilde{\Gamma}_{L2} \widetilde{\Gamma}_{R2}|G^R_{d22}(0)|^2 \right ),
\end{equation}
\begin{equation}
G^{DAR}=\frac{4e^2}{h}\widetilde{\Gamma}_{L1} \widetilde{\Gamma}_{L2}|G^R_{d12}(0)|^2,
\end{equation}
\begin{equation}
G^{CAR}=\frac{e^2}{h} \left ( \widetilde{\Gamma}_{L1} \widetilde{\Gamma}_{R2} +\widetilde{\Gamma}_{R1} \widetilde{\Gamma}_{L2} %%@
\right ) |G^R_{d12}(0)|^2 .
\end{equation}
It is obvious that all of the three different tunneling processes have contribution to the local conductance. Nevertheless the %%@
DAR tunneling process, as expected, has no contribution to the cross conductance. More interestingly, the CAR tunneling process %%@
provides a contrary contribution, in comparison with the ET process, to the cross conductance Eq.~(\ref{gc}), which is %%@
responsible for the negative value of the cross conductance in certain appropriate conditions as shown in the following section. %%@
This opposite role of the CAR can be interpreted in an intuitive way: a hole entering the right lead is physically equivalent to %%@
an electron injuring into the QD from the right lead, thus resulting in an opposite current flowing in the right lead. It is %%@
important to point out that if the supercoducting coupling is switched off ($\Gamma_s=0$), there are no DAR and CAR processes and %%@
as a result, the cross conductance reduces to the local conductance.

\section{Result And Discussion}

We suppose that the left and right leads are made from the identical material and in the wide band limit, of interest in the %%@
present investigation, the ferromagnetism of the leads can be accounted for by the polarization-dependent couplings $\Gamma_{L %%@
1}=\Gamma_{R 1}=(1+p)\Gamma$, $\Gamma_{L 2}=\Gamma_{R2}=(1-p)\Gamma$ for the parallel (P) alignment, while %%@
$\Gamma_{L1}=\Gamma_{R2}=(1+p)\Gamma$, $\Gamma_{L2}=\Gamma_{R1}=(1-p)\Gamma$ for the anti-parallel (AP) alignment. Here $\Gamma$ %%@
describes the tunneling coupling between the QD and the nonmagnetic leads, which is taken as the energy unit in the following %%@
calculations. And $p$ ($0\leq p< 1$) denotes the polarization strength of the leads.

In the following we will deal with the three-terminal QD system having a fixed finite Coulomb interaction $U=10$ at zero %%@
temperature and consider effects of changing bare dot level $\epsilon_d$ and the spin polarization $p$, and proximity strength %%@
$\Gamma_s$ respectively.

\subsection{Linear local and cross conductances}

\begin{figure}[htb]
\includegraphics[height=4.5cm,width=8.5cm]{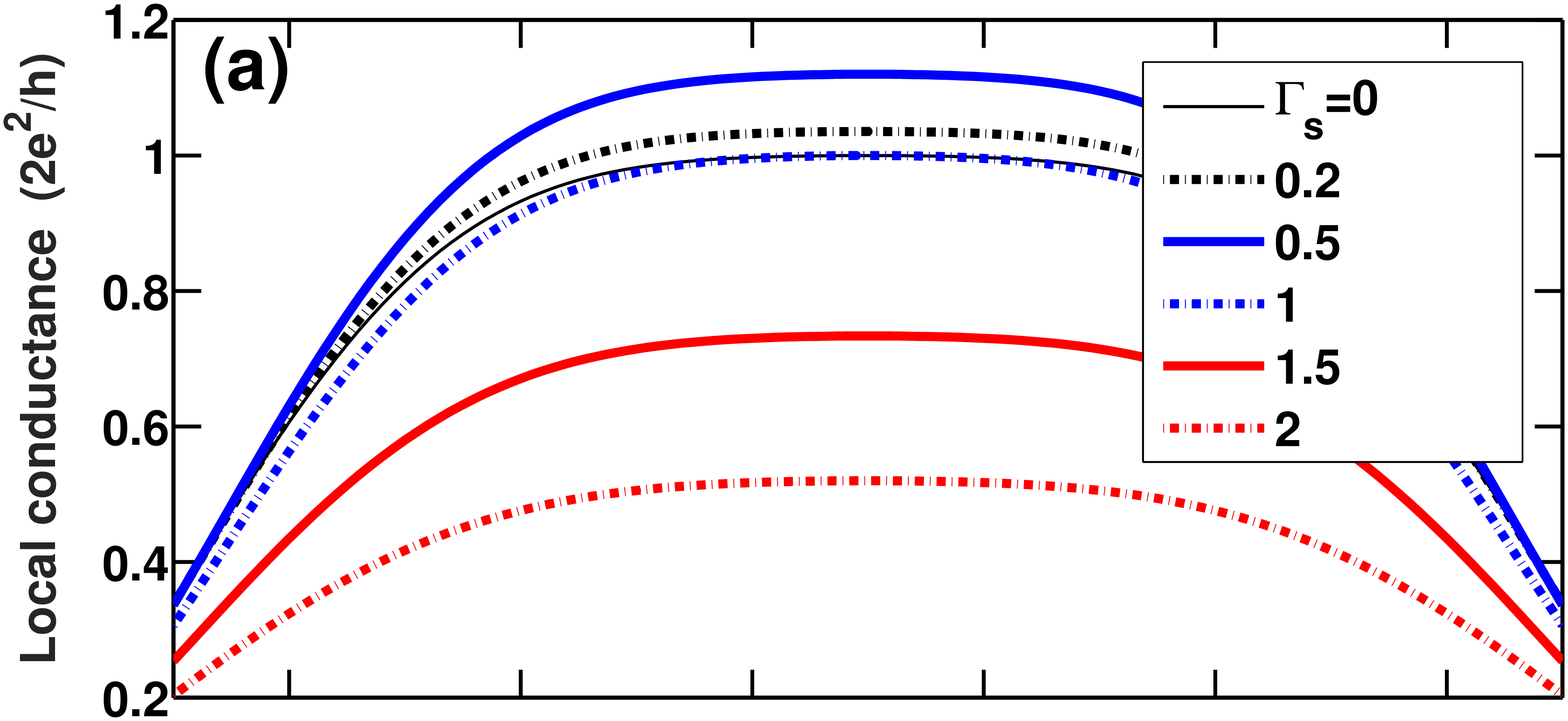}
\vspace{5mm}

\hspace{-1mm}\includegraphics[height=5cm,width=8.5cm]{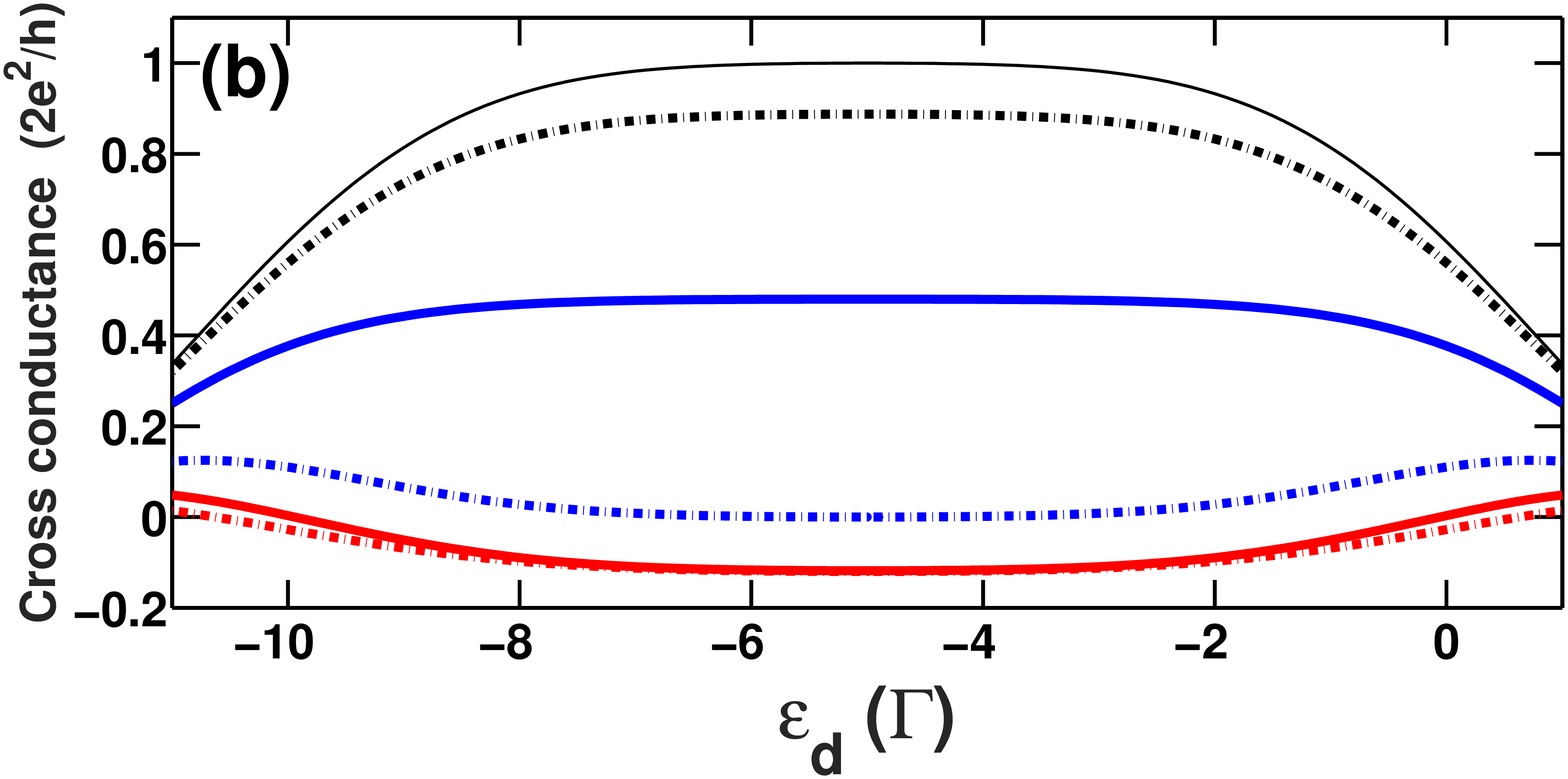}
\caption{(Colour online) (a) The local conductance and (b) the cross conductance versus the bare dot level $\epsilon_d$ for %%@
different proximity-coupling strengths $\Gamma_s$ in the case of normal leads, i.e., $p=0$.}
\label{fig2}
\end{figure}

We show, at first, the calculated linear conductances in Fig.~2, including the local conductance $G_L$ and the nonlocal cross %%@
conductance $G_C$ as functions of the bare energy level $\epsilon_d$ of the QD at different superconducting coupling strengths %%@
$\Gamma_s=0$, $0.2$, $0.5$, $1.0$, $1.5$, and $2.0$ in the case of no spin-polarization $p=0$. Without the superconducting %%@
coupling $\Gamma_s=0$, $G_L=G_C$ and the linear conductance reaches the unitary limit, $G_0$ ($G_0\equiv 2e^2/h$), as expected in %%@
the Kondo regime. With increasing the coupling $\Gamma_s$, the local conductance $G_L$ raises at the beginning, as seen in %%@
Fig.~2(a), since the AR channel starts to emerge and make contribution to the electronic tunneling. A little bit bigger value of %%@
the conductance, $G_L\simeq 1.1 G_0$, than the unitary limit of conductance of single-particle tunneling is reached at the %%@
coupling $\Gamma_s=0.5$ in the Kondo regime.
It is important to notice that such bigger value of the conductance is a signature indicating that the tunneling event in the %%@
present hybrid system is mixture of the single-particle and Cooper pair tunnelings.
Increasing furthering the coupling $\Gamma_s$ will, however, cause decrease of the local conductance $G_L$. The suppression of %%@
$G_L$ can be interpreted as follows: electron coming from the left lead has much more higher probability to form the Cooper pair %%@
injuring into the superconducting electrode due to the considerably strong coupling $\Gamma_s>0.5\Gamma$, and as a result, the ET %%@
process is rapidly suppressed. Different from the local conductance, the nonlocal conductance $G_C$ decreases from the beginning %%@
and even becomes negative if the proximity-coupling is sufficiently strong.
The negative cross conductance means that, when the left lead is applied voltage bigger than the right lead, electrons will, %%@
instead of entering into the right lead from the QD, tunnel into the QD out of the right lead. Moreover, we find that when the QD %%@
leaves the Kondo regime, the cross conductance will become positive again.

Such effects of $\Gamma_s$ are clearly manifested in Fig.~3, in which the local and nonlocal conductances, and their three %%@
respective parts, $G^{ET}$, $G^{DAR}$, and $G^{CAR}$ as well, are illustrated as functions of the coupling $\Gamma_s$ for the %%@
specific system having bare dot level, $\epsilon_d=-U/2=-5$. It is observed that: a maximum value of the local conductance, %%@
$G_L=1.125G_0$ is arrived at $\Gamma_s=0.58\Gamma$; After that point of $\Gamma_s$, the AR process becomes the predominate %%@
tunneling mechanism over the ET process; When the proximity-coupling is equal to the tunnel-coupling, i.e. $\Gamma_s=\Gamma$, a %%@
new resonance is reached originated from interplay between the Kondo effect and AR. As consequences,  $G^{DAR}=G_0/2$ and %%@
$G^{CAR}=G^{ET}=G_0/4$, and the local conductance is thus arrived at the unitary value, $G_L=G_0$ again. At the same time, the %%@
nonlocal conductance is completely vanished, $G_C=0$, which indicates no current response in the right lead to the bias voltage %%@
applied to the left lead.

\begin{figure}[htb]
\includegraphics[height=5.5cm,width=8.5cm]{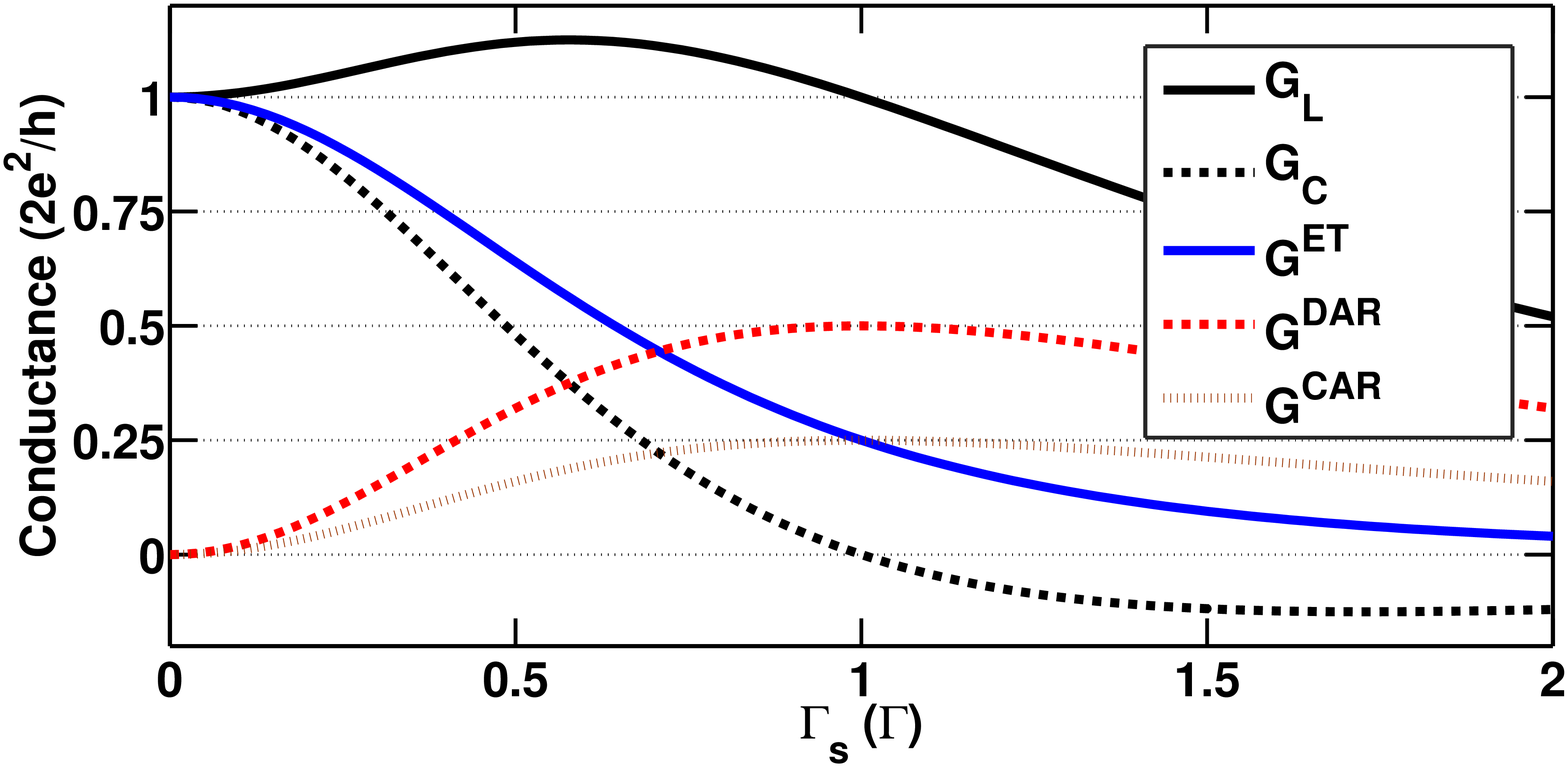}
\caption{(Colour online) The local conductance (black-solid line) and the cross conductance (black-dotted line) versus the %%@
proximity coupling $\Gamma_s$ for the system with a bare dot level at the particle-hole symmetric point, $\epsilon_d=-U/2=-5$ in %%@
the case of normal leads. The three parts of the conductance are also plotted as well for illustration.}
\label{fig3}
\end{figure}

\begin{figure}[htb]
\includegraphics[height=4.5cm,width=8.5cm]{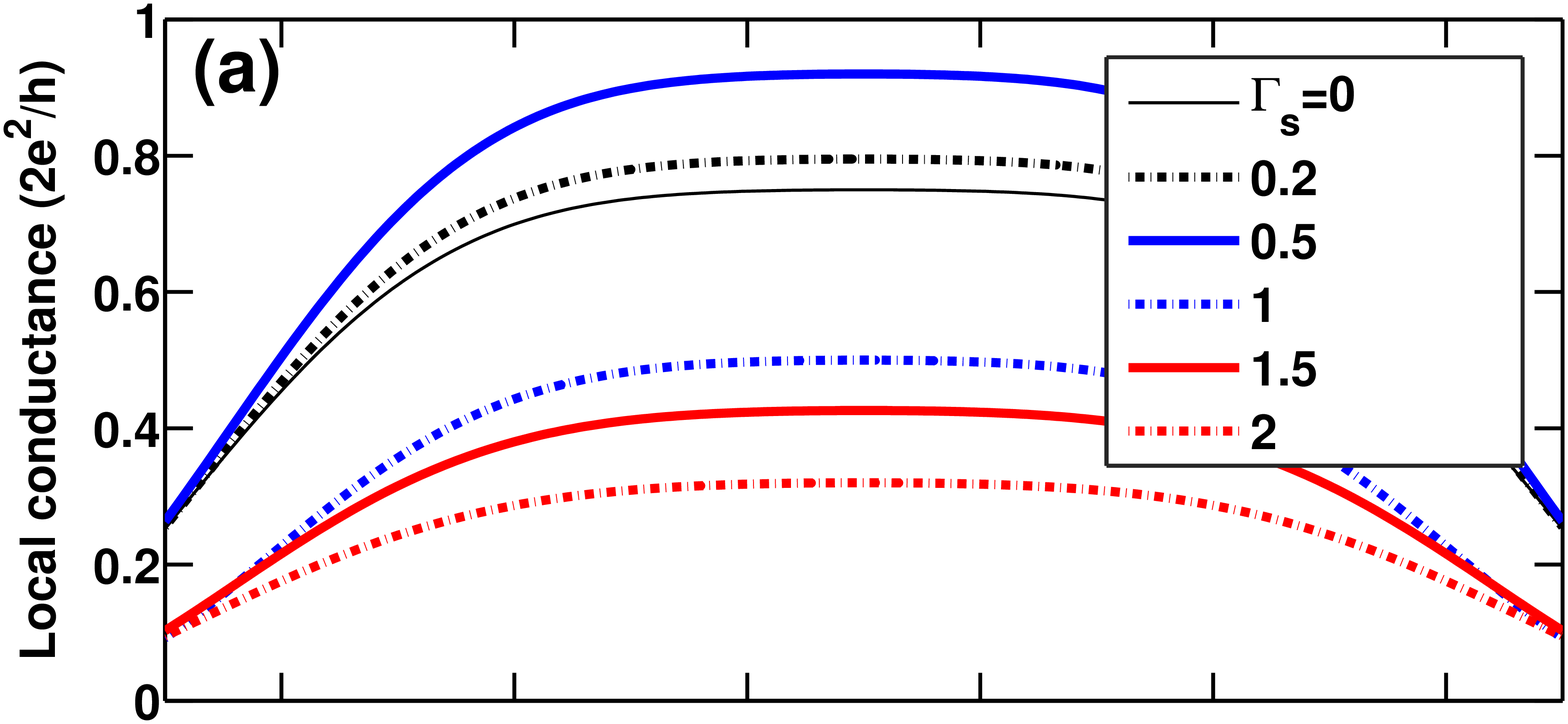}
\vspace{5mm}

\hspace{-1.5mm}\includegraphics[height=5cm,width=8.5cm]{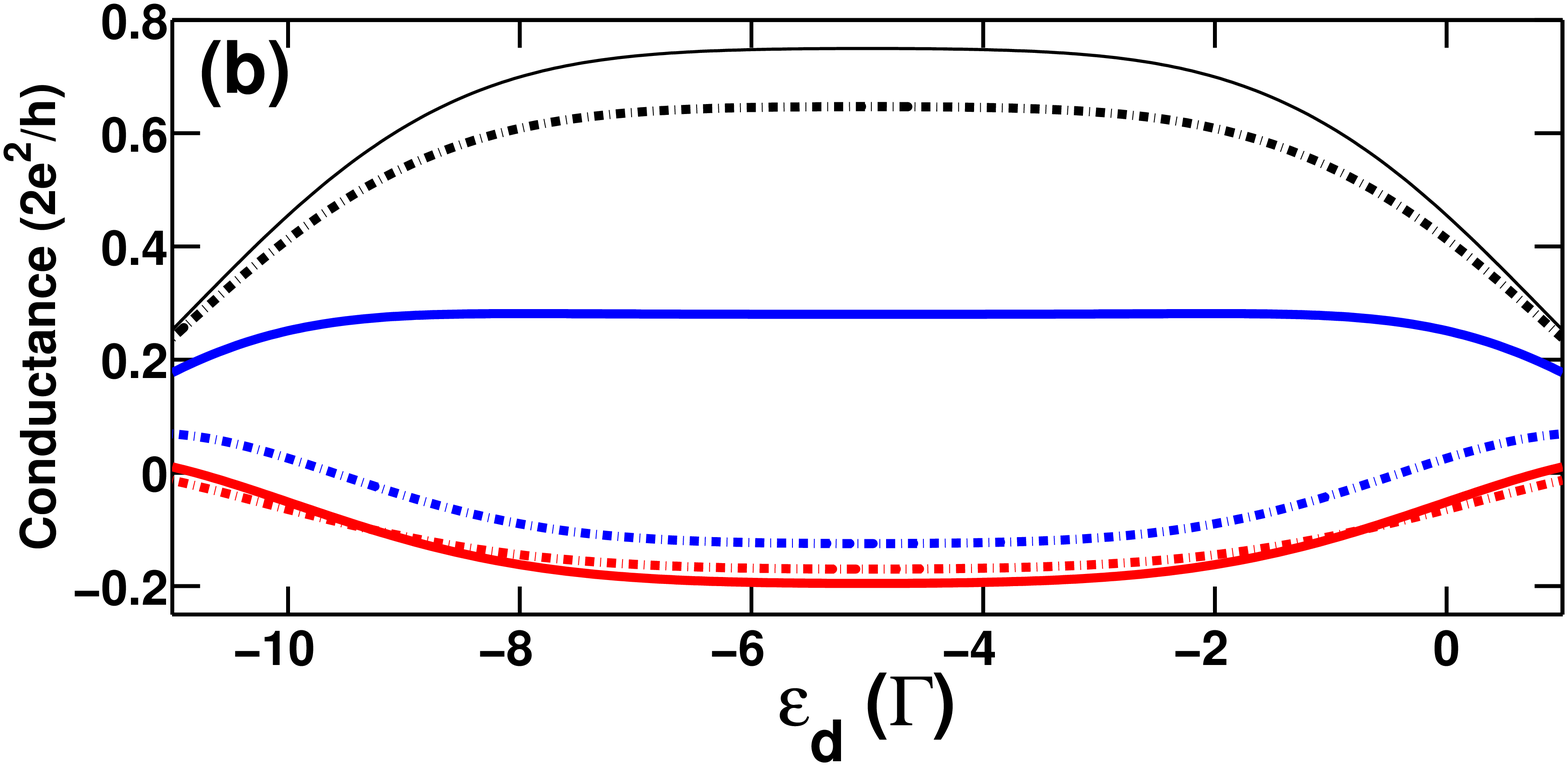}
\caption{(Colour online) (a) The local conductance and (b) the cross conductance versus the bare dot level $\epsilon_d$ for %%@
different proximity-coupling strengths $\Gamma_s$ in the AP configuration with $p=0.5$.}
\label{fig4}
\end{figure}

Secondly, we investigate the cross conductance $G_C$ as a function of the bare energy level $\epsilon_d$ of the QD at different %%@
proximity couplings $\Gamma_s$ in the AP configuration with a large spin polarization $p=0.5$ in Fig.~\ref{fig4}. In the AP %%@
configuration, similar with the case of zero spin polarization $p=0$, electrons with up-spin and down-spin are equally available %%@
in the whole system, favoring the formation of the Kondo-correlated state within a wide dot level range centered at %%@
$\epsilon_d=-U/2=-5$. Meanwhile, since there is no splitting of the renormalized dot levels, $\epsilon_d+ \lambda_\sigma^2$, for %%@
different spins, the usual tunneling and charging peaks, around $\epsilon_d=0$ and $-U$ respectively, are relatively narrow. The %%@
local conductance $G_L$-vs.-$\epsilon_d$ curves show similar behavior with the case of zero spin polarization even in the %%@
presence of superconducting coupling $\Gamma_s$. Furthermore, since no spin-flip scattering exists in the tunneling processes, in %%@
the AP configuration the majority-spin (e.g. up-spin) states in the left lead increase but the available up-spin (minority-spin) %%@
states in the right lead decrease with increasing spin polarization strength, and as a consequence the transfer of the %%@
majority-spin (up-spin) electrons through the QD is suppressed, such that the local conductance goes down and eventually vanishes %%@
at $p=1$ as expected. On the contrary, the available down-spin states in the right lead indeed increase in the AP configuration, %%@
which just facilitates occurrence of the CAR process.\cite{PhysRevLett.93.197003} Therefore, one can observe that $G_C$ becomes %%@
negative at almost the whole region of dot levels, from the mixed-valence regime to the empty orbital regime even when $\Gamma_s< %%@
1$, and arrives nearly at a considerably bigger negative value, $G_C\simeq -G_0/5$, at the Kondo regime at $p=0.5$.
It is physically interesting to consider the extreme case of $p=1$. As mentioned above, in the AP configuration electrons with %%@
up-spin and down-spin are identical with each other, preferring the formation of the Kondo-correlated state for all values of %%@
$p$. However, since the up-spin states are almost unavailable in the right lead in the case of large polarization, the ET process %%@
for the left lead to the right lead is completely damaged (implying an exactly vanishing conductance in the usual QD system), but %%@
the CAR process survives here as unique tunneling mechanism and makes contribution to electronic tunneling exclusively. It is %%@
anticipated that in this case, $G^{ET}=G^{DAR}=0$ and $G_L=-G_C=G^{CAR}=G_0/2$ (this is the unitary limit of conductance of %%@
single channel).

\begin{figure}[htb]
\includegraphics[height=4.5cm,width=8.5cm]{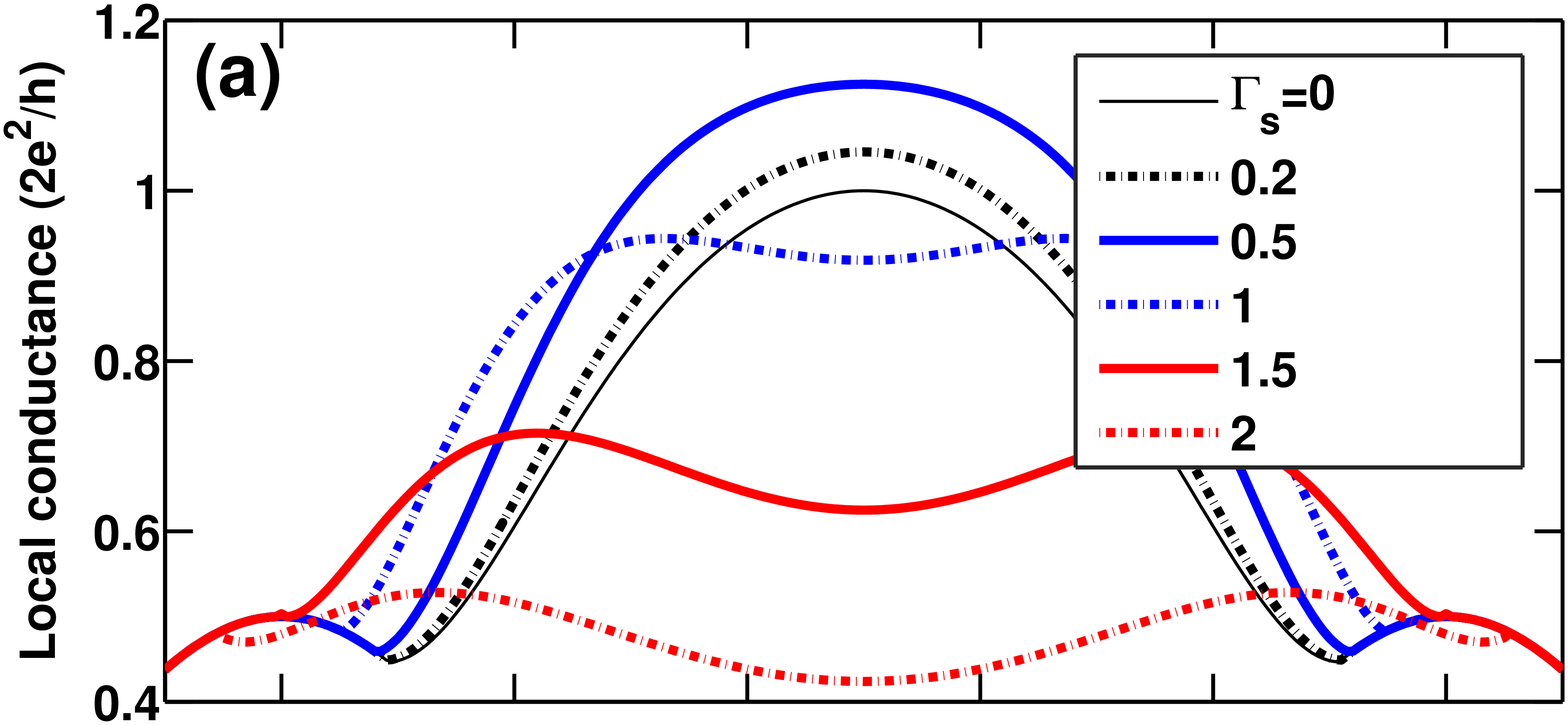}
\vspace{5mm}

\hspace{-1.5mm}\includegraphics[height=5cm,width=8.5cm]{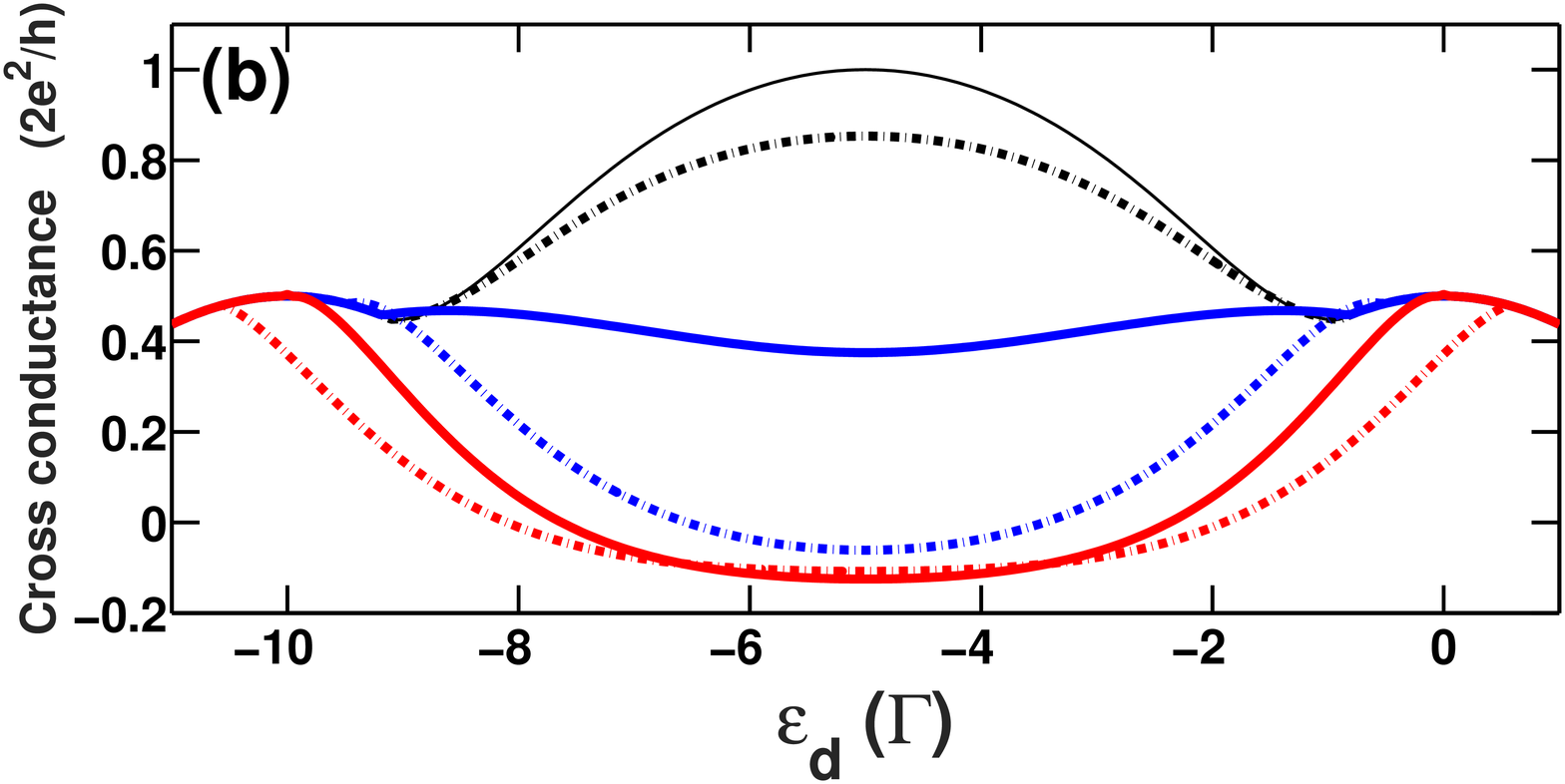}
\caption{(Colour online) (a) The local conductance and (b) the cross conductance versus the bare dot level $\epsilon_d$ for %%@
different proximity-coupling strengths $\Gamma_s$ in the P configuration with $p=0.5$.}
\label{fig5}
\end{figure}

The situation is quite different in the case of P configuration, as demonstrated in Fig.~\ref{fig5}, in which the two %%@
conductances are plotted as functions of bare dot level with spin polarization $p=0.5$. In the P configuration, finite spin %%@
polarization splits the dot level for up and down spins and thus broadens the usual resonance peaks around $\epsilon_d=0$ and %%@
$\epsilon_d=-U$.\cite{PhysRevLett.91.127203,Pasupathy86,hauptmann2008,PhysRevLett.107.176808,Dong_2003} On one hand, since %%@
minority-spin electrons are still available in the two electrodes to build Kondo screening correlation to certain degree, the %%@
central Kondo peak can still be reached at the unitary limit $G_0$ at the large polarization $p=0.5$ in the case of $\Gamma_s=0$. %%@
On the other hand, the number of minority-spin electrons is to small to construct the Kondo-correlated state at $p=0.5$ and thus %%@
Kondo-induced conductance enhancement disappears rapidly when the QD leaves away from the particle-hole symmetric point %%@
$\epsilon_d=-U/2$.
These two factors cause the appearance of kinks or splitting peaks in the both conductance-vs.-$\epsilon_d$ curves.
Besides, it is observed from Fig.~\ref{fig5}(a) that the central Kondo peak in the local conductance is also progressively %%@
splitting with increasing proximity coupling $\Gamma_s\geq \Gamma$ in this P configuration.
Furthermore, decrease of minority-spin states in both leads in the P configuration hinders emergence of AR processes, which leads %%@
to weakly negative cross conductance in the Kondo regime, e.g. $G_C\geq -0.1G_0$, and even totally vanished CAR, thus %%@
$G_{C}\simeq G_{L}$, at the two usual resonance peaks as shown in Fig.~\ref{fig5}(b). It states that strong ferromagnetism %%@
destroys proximitized superconductivity in this three-terminal hybrid nanosystem.

\subsection{Nonlinear local and cross Conductances}

Now we turn to investigation of nonlinear tunneling, since the nonlinear differential conductance $dI_L/dV$ is believed to be a %%@
very useful tool in experiments to detect the formation of the Kondo-correlated state due to its proportionality
to transmission spectrum, supposed that the total transmission is unchanged subject to external bias voltage. In the present %%@
three-terminal hybrid device, one can define the local and cross differential conductances, $g_L=\partial I_L/\partial V$ and %%@
$g_C=\partial I_R/\partial V$, if the bias voltage $V$ is applied to the left lead and meanwhile the superconductor and the right %%@
lead are kept grounded. From the current formulas Eqs.~(\ref{iL})-(\ref{icar}), we can then obtain that the two diffenertial %%@
conductances are both proportional to the normal transmission spectrum $T_N(\omega)$ and the AR spectrum $T_A(\omega)$ at %%@
$\omega=V$ at zero temperature, $g_L\propto T_N(V)+aT_A(V)$ and $g_C\propto T_N(V)-bT_A(V)$ ($a$ and $b$ are constants).

\begin{figure}[htb]
\includegraphics[height=4.5cm,width=8.5cm]{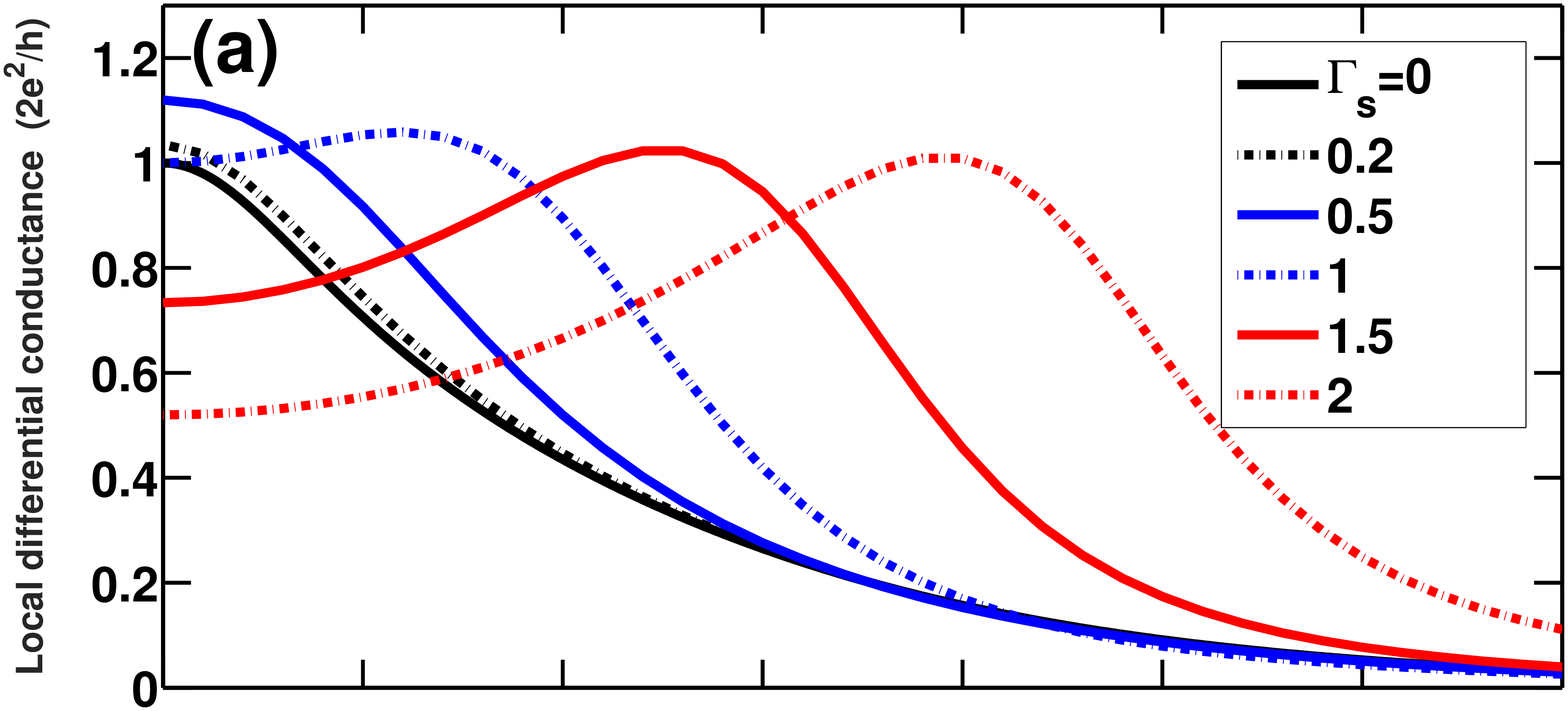}
\vspace{5mm}

\hspace{-1mm}\includegraphics[height=5cm,width=8.5cm]{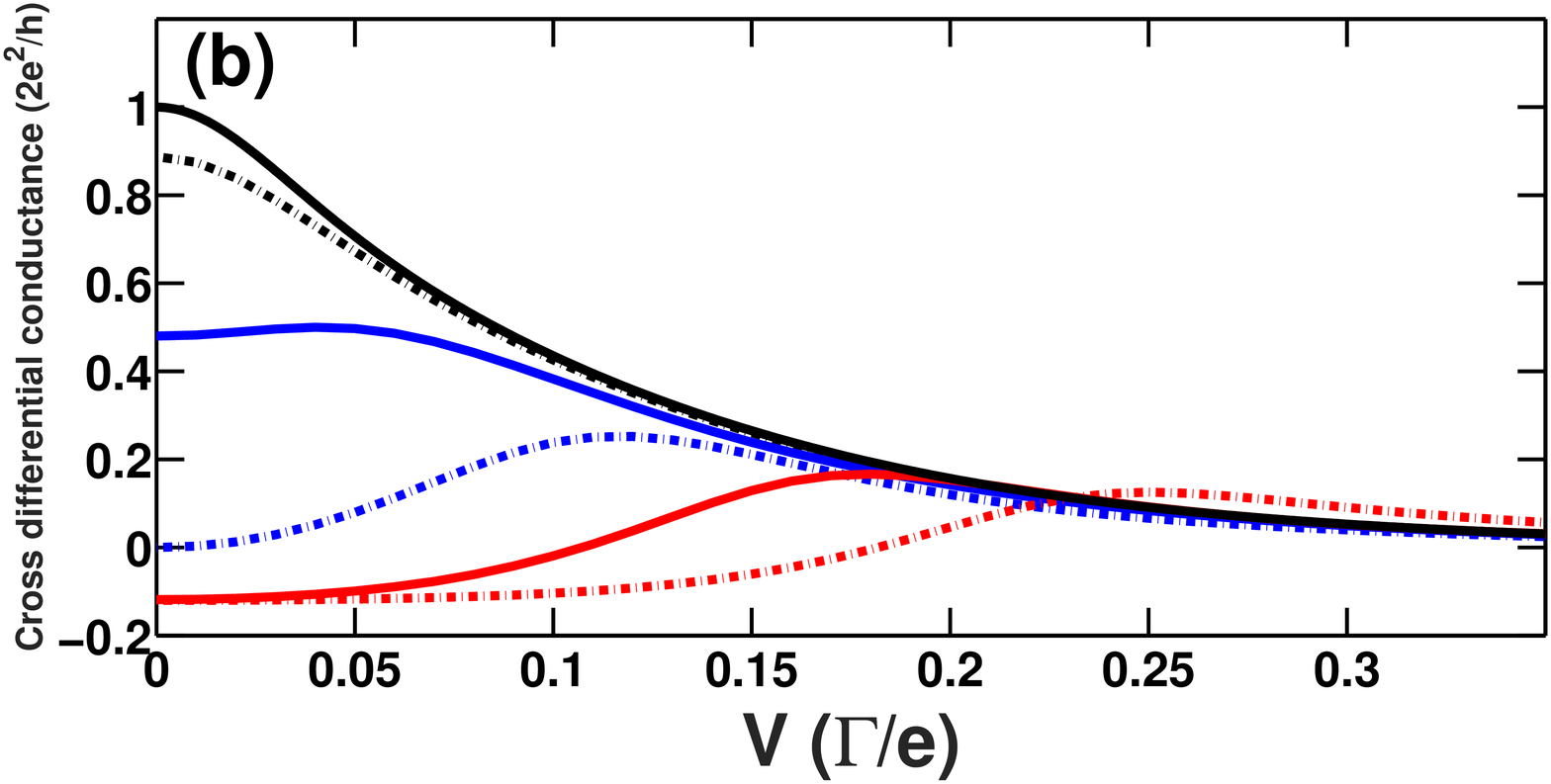}
\caption{(Colour online) The zero-temperature local (a) and cross (b) differential conductances versus bias voltage $V$ for %%@
various couplings $\Gamma_s$ for the system with bare dot level $\epsilon_d=-5$ in the case of normal leads ($p=0$).}
\label{fig6}
\end{figure}

Figure \ref{fig6} shows the local and cross differential conductances as functions of bias voltage at various proximity couplings %%@
$\Gamma_s$ for the system having a single dot level $\epsilon_d=-5$ at the Kondo regime.
These curves for weak proximity coupling $\Gamma_s<\Gamma$ present a single zero-bias anomaly, being the signature of the Kondo %%@
effect. Nevertheless, there appears nonzero-bias peak with increasing proximity coupling $\Gamma_s\geq \Gamma$. It is announced %%@
that the Kondo correlation enhances not only the normal ET but also the AR, nonetheless the increasing superconducting proximity %%@
coupling induces splitting of the Kondo peaks in the normal transmission spectrum and the AR spectrum as well. This peak %%@
splitting is the reason that three parts of the linear conductance are all suppressed when $\Gamma_s>\Gamma$ as shown in %%@
Fig.~\ref{fig3}. Finally, one can observe that the negative cross differential conductance becomes positive at the case of large %%@
bias voltage. External bias voltage plays a role of dissipation so as to destroy not only the Kondo correlation but the negative %%@
nonlocal current response as well.

\section{Conclusion}

We have theoretically investigated the subgap transport properties of a hybrid nanosystem consisting of an interacting QD %%@
connected to one superconducting lead and two ferromagnetic leads.
Based on finite-$U$ slave boson mean field approach and NGF method, we find markedly rich transport features ascribed to the %%@
competition among the Kondo correlation, superconducting proximity effect, and spin polarization of electrodes. In the case of %%@
weak superconducting proximity coupling, the Kondo-correlated state can still be built leading to a single zero-bias peak in the %%@
voltage-dependent differential conductance. But the peak height drops down gradually with increasing $\Gamma_s$, and when %%@
$\Gamma_s\geq \Gamma$, a non-zero peak appears. Such strong proximity coupling induces linear cross conductance negative in the %%@
Kondo region. Spin polarization can further enhance opposite current response in the right lead, i.e. more negative cross %%@
conductance, in the AP configuration, because such configuration is advantageous to the emergence of CAR. In contrast, in the P %%@
configuration, rising spin polarization $p$ blocks CAR process and also splits the Kondo peak, such that the linear local %%@
conductance exhibits four peaks behavior when $\Gamma_s\geq \Gamma$, and the linear cross conductance reduces to the normal %%@
positive conductance more rapidly.

\begin{acknowledgments}

This work was supported by Projects of the National Science Foundation of China under Grant No. 11674223.

\end{acknowledgments}

\bibliography{sqdbib}

\end{document}